\documentclass[reprint,sort&compress,longbibliography,superscriptaddress,amsmath,aps,showpacs,pra]{revtex4-2}
\usepackage[british]{babel}
\usepackage{graphicx,epsfig}
\usepackage{float}
\usepackage[hidelinks]{hyperref}
\usepackage{breqn}
\usepackage[utf8]{inputenc}
\usepackage{amsmath}
\usepackage{subfigure}
\usepackage[normalem]{ulem}
\usepackage{xcolor}
\usepackage{hyperref}
\usepackage{soul}
\hypersetup{ 
     colorlinks   = true,
     citecolor    = blue,
}

\begin{document}
\title{High transmission in twisted bilayer graphene with angle disorder}
\author{H\'ector Sainz-Cruz}
\affiliation{$Imdea\ Nanoscience,\ Faraday\ 9,\ 28015\ Madrid,\ Spain$}
\author{Tommaso Cea}
\affiliation{$Imdea\ Nanoscience,\ Faraday\ 9,\ 28015\ Madrid,\ Spain$}
\author{Pierre A. Pantale\'on}
\affiliation{$Imdea\ Nanoscience,\ Faraday\ 9,\ 28015\ Madrid,\ Spain$}
\author{Francisco Guinea}
\affiliation{$Imdea\ Nanoscience,\ Faraday\ 9,\ 28015\ Madrid,\ Spain$}
\affiliation{$Donostia\ International\  Physics\ Center,\ Paseo\ Manuel\ de\ Lardizabal\ 4,\ 20018\ San\ Sebastian,\ Spain$}
\affiliation{$Ikerbasque\  Foundation, Maria\ de\ Haro\ 3,\ 48013\ Bilbao,\ Spain$}
\date{\today}
\begin{abstract}
 Angle disorder is an intrinsic feature of twisted bilayer graphene and other moir\'e materials. Here, we discuss electron transport in twisted bilayer graphene in the presence of angle disorder. We compute the local density of states and the Landauer-B\"uttiker transmission through an angle disorder barrier of width comparable to the moir\'e period, using a decimation technique based on a real space description. We find that barriers which separate regions where the width of the bands differ by 50$\%$ or more lead to a minor suppression of the transmission, and that the transmission is close to one for normal incidence, which is reminiscent of Klein tunneling. These results suggest that transport in twisted bilayer graphene is weakly affected by twist angle disorder.
  
\end{abstract}
\maketitle

{\it Introduction.} Twisted bilayer graphene (TBG) is a superconductor \cite{cao2018unconventional, yankowitz2019, lu2019superconductors} and shows many more correlated phases \cite{li2010, cao2018correlated, xie2019, kerelsky2019, jiang2019charge, choi2019, polshyn2019large, sharpe2019emergent, cao2020strange, serlin2020intrinsic, chen2020tunable, saito2020independent, zondiner2020cascade, wong2020cascade, stepanov2020untying, arora2020superconductivity, xu2020correlated, nuckolls2020strongly, saito2021hofstadter, choi2021correlation, park2021flavour, liu2021tuning, rozen2021entropic, saito2021isospin, cao2021nematicity, stepanov2020competing, pierce2021unconventional}. The microscopic origins of superconductivity and almost all other phases are still unknown and intensely debated \cite{lopes2007graphene, morell2010flat, bistritzer2011moire, sanjose2012nonabelian, nam2017lattice, po2018fragile, po2018origin, kang2018symmetry, koshino2018maximally, wu2018theory, isobe2018unconventional, guinea2018electrostatic, gonzalez2019kohn, tarnopolsky2019origin, wu2019phonon, angeli2019valley, lewandowski2019intrinsically, chichinadze2020nematic, fernandes2020nematicity, bultinck2020ground, phong20obstruction, xie2021twisted, gonccalves2020incommensurability, khalaf2020soft}. Recent highlights in the field are the discovery of superconductivity in mirror-symmetric twisted trilayer graphene \cite{park2021tunable, hao2021electric} and rhombohedral trilayer graphene \cite{zhou2021superconductivity}, as well as the creation of Josephson junctions made of different regions of single TBG crystals \cite{de2021gate, rodan2021highly}. The phase diagram of TBG strongly depends on the twist angle, which is a novel tuning parameter but also an unavoidable source of disorder that creates a unique landscape of domains with different angles in each sample. In devices that show superconductivity and insulating behavior, the angle disorder can be of order $0.1^{\circ}$ over micrometers \cite{uri2020}. On the other hand, certain superconducting domes and insulating phases have been observed in samples with very low angle disorder ($\lesssim0.02^{\circ}$ throughout the device) \cite{lu2019superconductors, stepanov2020competing}. Angle disorder is a natural candidate to account for part of the variability in phase diagrams, since it modulates the interlayer hopping, the Fermi velocity and the bandwidth.

Early experiments on TBG observed moir\'e patterns of different periods through scanning tunneling microscopy (STM) \cite{li2010,luican2011, brihuega2012, wong2015} and several groups have used STM to measure local twist angles from distortions in the moir\'e structure \cite{kerelsky2019, choi2019, xie2019, jiang2019charge}. In parallel, transmission electron microscopy has enabled the observation of lattice reconstruction \cite{yoo2019, kazmierczak2021strain}. Recently, Raman spectroscopy has emerged as a valuable probe of angle disorder \cite{schapers2021raman} and lattice dynamics \cite{gadelha2021localization}. 
Angle disorder was precisely characterized by Uri \textit{et. al.} \cite{uri2020} who use a SQUID to measure quantum Hall currents stemming from the Landau level structure and which occur in the bulk of the sample across twist-angle equi-contours.
The angle maps of two superconducting samples reveal the coexistence of smooth angle gradients together with stacking faults and dislocations. Lattice distortions also occur due to strain, which is concomitant with long range angle disorder. However, periodic tensions between layers develop always and result in constant deformations of the unit cell, independent of disorder. In particular, uniaxial heterostrain has a major impact on the physics of TBG near the magic angle \cite{bi2019designing, mesple2020heterostrain, parker2021strain}. Together with angle disorder and strain, TBG also has moderate charge disorder \cite{tilak2020}.

There are few theoretical works addressing angle disorder in TBG \cite{wilson2020, padhi2020, joy2020transparent, thomson2021recovery}. In Ref. \cite{wilson2020}, the effect of angle disorder domains on the densities of states (DOS) was studied. The bandwidth of the flat bands, the gaps to remote bands and the sharpness of the van-Hove peaks were found to be sensitive to angle disorder. So far, transport in the presence of angle disorder has only been studied with a two-band version of the continuum model \cite{padhi2020} and focused on tunneling through narrow angle domain walls. Another paper \cite{joy2020transparent} describes the transmission of Dirac particles through quasi-1D angle domains, modelled as a change in the Fermi velocity plus a gauge field that shifts the Dirac points. In Ref. \cite{thomson2021recovery}, a Landau-Ginzburg theory of strong correlations coupled to angle disorder is developed in an attempt to explain the appearance of semi-metallic phases near charge neutrality.

In this paper, we study electron transport in TBG across finite-width angle disorder barriers using a tight binding model. To avoid the large moir\'e unit cells associated to small twist angles, we perform a scaling approximation \cite{gonzalez2017, vahedi2021magnetism}. The transmission through the angle disorder barrier is calculated in the Landauer-B\"uttiker formalism by means of a decimation technique that obtains the dressed Green's function at the barrier. We find that for disorder of the magnitude seen in some superconducting samples ($\sim0.1^{\circ}$), the transmission through an angle disorder barrier is high, while the density of states at the barrier shows van-Hove peaks significantly broadened. Although conventional Klein tunneling \cite{katsnelson2006} cannot occur in our system, the transmission is complete ($>99\%$) for normal incidence. 
The transmission within energy windows mixing ten channels, normal and oblique, depends linearly on the bandwidth ratio of the angles on either side of the barrier.

\begin{figure}
\hspace*{-0.25cm}
    \centering
    {\includegraphics[width=8cm]{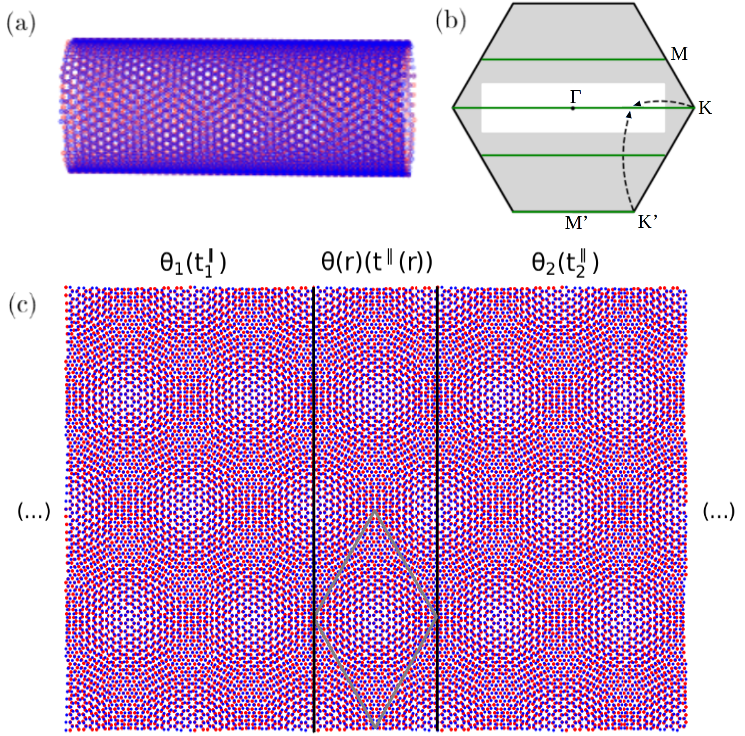}}%
    \caption{(a) Sketch of the crystal lattice of a TBG nanotube. (b) Folding the mini-Brilluoin zone (mBZ) of TBG onto the white rectangle and superimposing the bands along the green lines, one obtains the 1D band structure of the TBG nanotube. (c) Setup of transmission calculation. The section between vertical lines is the planar projection of the unit cell of TBG nanotube (56,2)@(-56-2), with N=4336 sites, where intralayer hopping disorder $t^{\parallel}(r)$ mimics angle disorder $\theta(r)$ and which connects to two semi-infinite pristine regions or `leads'. The rhombus shows a unit cell of planar TBG.}
    \label{fig:1}
\end{figure}

{\it Model.} The system we study is a double-wall carbon nanotube made of two nanotubes with opposite chiral angles, the 1D analog of TBG, see Fig. \ref{fig:1}(a). Such a system is entirely defined by the chiral vectors of its constituent nanotubes $\vec{C}_\pm=\pm(n\vec{a}_1+m\vec{a}_2)\equiv\pm(n,m)$, where $\vec{a}_{1,2}=(\sqrt{3}/2,\pm1/2)a$ are basis vectors of graphene and $a=0.246$ nm is the lattice constant. The chiral vector specifies a possible wrapping of the graphene lattice into a tube; it also determines the chiral angle, which is equivalent to a twist angle and corresponds to the tilt angle of the honeycomb lattice with respect to the tube's axis. Imposing periodic boundary conditions to close the nanotube quantizes the momentum parallel to the barrier, and the only allowed values correspond to the green lines in Fig. \ref{fig:1}(b). We will call the system a `TBG nanotube', and denote it as (n,m)@(-n,-m). Figure \ref{fig:1} shows a section of TBG nanotube (56,2)@(-56,-2), which has a relative chiral angle (i.e. twist angle) of 3.48$^{\circ}$. The central unit cell between lines is the angle disorder barrier that connects to semi-infinite pristine leads. We focus on the large diameter limit, in which the TBG nanotube reproduces the spectrum of TBG, just like a single-wall nanotube becomes equivalent to graphene in that limit. The low diameter limit, in which the 1D nature takes center stage, has recently been explored and shown to enable strong interlayer coupling \cite{koshino2015incommensurate, zhao2020}. Here, the main advantages in the use of nanotubes are the knowledge of carriers' momenta parallel to the barrier and the avoidance of border effects. To model electronic properties, we use a `minimum' tight binding Hamiltonian \cite{lin2018minimum}, 

\begin{dmath}
{H=H_{\parallel}+H_{\bot}+H_{dip}=-\sum_{i\neq j,m}\gamma_{ij}^{mm}(c^{\dagger}_{i,m}c_{j,m}+h.c.)}-\sum_{i, j,m}\gamma_{ij}^{m,m+1}(c^{\dagger}_{i,m}c_{j,m+1}+h.c.)+\sum_{i,m}V_{dip}(r)c^{\dagger}_{i,m}c_{i,m}\, .
\label{eq:1}
\end{dmath}

Here $i,j$ run over the lattice sites and $m$ is the layer index. The first term in the Hamiltonian is an intralayer hopping to nearest-neighbours only and the second an interlayer hopping that decays exponentially away from the vertical direction,

\begin{equation}
\gamma_{ij}^{mm}=t_{\parallel}\, ,
\label{eq:2}
\end{equation}
\begin{equation}
\gamma_{ij}^{m,m+1}=t_{\perp}e^{-(\sqrt{r^2+d_0^2}-d_0)/\lambda_\perp)}\frac{d_0^2}{r^2+d_0^2}\, ,
\label{eq:3}
\end{equation}
where $d_0=0.335$ nm is the distance between layers, $t_{\parallel}=3.09$ eV and $t_{\perp}=0.39$ eV are the intralayer and interlayer hopping amplitudes and $\lambda_\perp=0.027$ nm is a cutoff for the interlayer hopping \cite{lin2018minimum}. This model captures all the essential features of the band structure of magic-angle TBG, i.e. flat bands separated by gaps from remote bands, van-Hove singularities near the M points and Dirac cones near K and K' points. 

In experiments, one imposes a nearly uniform electron density through the back-gate voltage. In regions with narrower bands this density translates into a higher chemical potential. To avoid unbalanced charge, carriers redistribute near the angle barrier and create a dipole potential $V_{dip}$ that compensates the chemical potential mismatch, thus balancing the Fermi energies on either side of the barrier. This leads to large unscreeened in-plane electric fields \cite{uri2020}. The chemical potential difference corresponds to the offset in Dirac point energies obtained in the band structure calculation, and is $\Delta_\mu=3.67$ meV in case of 1.11$^{\circ}$ coupled to 1.21$^{\circ}$. Thus, at this simplest level of description, $V_{dip}(r)$ is an on-site potential that evolves smoothly as an hyperbolic tangent from 0 to $\Delta_\mu$ only within the barrier, analogous to the function that introduces angle disorder $\theta(r)$. A more rigorous dipole potential would extend far into the pristine leads, but its self-consistent calculation is beyond the scope of this research.

\begin{figure}
    \centering
    {\includegraphics[scale=0.55]{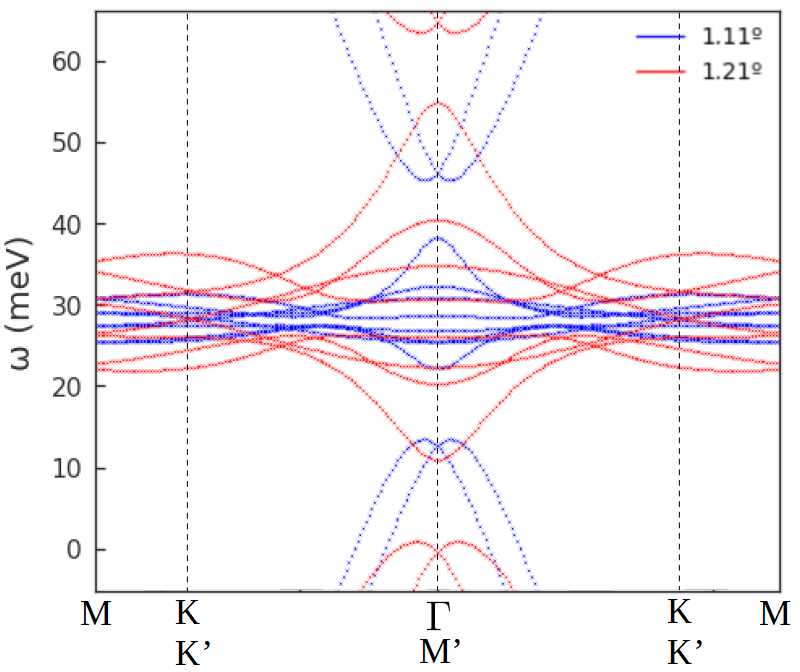}}%
    \caption{Low-energy band structures of TBG at 1.11$^{\circ}$ and 1.21$^{\circ}$. The bands are obtained through exact diagonalization of a scaled tight binding model of TBG nanotube (56,2)@(-56,-2) ($\theta=3.48^{\circ}$).}
    \label{fig:2}
\end{figure}

Using the continuum model, it can be shown that the bands of twisted bilayer graphene depend, to first order, on a dimensionless parameter \cite{bistritzer2011moire}, which involves the ratio of interlayer to intralayer hopping and the twist angle,

\begin{equation}
\alpha = \frac{at_{\perp}}{2\hbar v_F \sin ( \theta / 2 )} \propto \frac{t_{\perp}}{t_{\parallel}\theta}\, .
\label{eq:4}
\end{equation}
where $v_F$ is the Fermi velocity and $t_\perp$ is an average of the interlayer hopping. We emphasise two important consequences. First, this relation enables simulation of smaller angles with an appropriate reduction in $t_{\parallel}$ \cite{gonzalez2017,vahedi2021magnetism, SM}. Second, within this one-parameter model, there is an equivalence between angle disorder $\theta(r)$ and intralayer hopping disorder $t_{\parallel}(r)$ \cite{SM}, since both result in identical $\alpha(r)$-disorder. As a consequence, both angle disorder and intralayer hopping disorder capture very well the pronounced changes in bandwidth near the magic angle. In virtue of this, we have approximated angle disorder as intralayer hopping disorder in the transmission calculation, which has the advantage of preserving translational symmetry. Nevertheless, hopping disorder does not reflect the small modifications in moir\'e period that angle disorder would also provoke. Therefore, the approximation rests on the assumption that, near the magic angle, moir\'e period variations due to a twist angle difference of 0.1$^{\circ}$ (typically $\lesssim 10\%$) have a negligible impact on transport compared to the more drastic bandwidth changes ($\gtrsim 150\%$ or more, see Figure \ref{fig:2}). 

To study transport, we consider the standard setup of a conductor coupled to semi-infinite leads \cite{datta1997electronic}. We compute the transmission through an angle disorder barrier of width one moir\'e unit cell (as in Figure \ref{fig:1}) or two, connected to left and right to semi-infinite leads which are pristine in twist angle. The transmission is calculated in the Landauer-B\"uttiker formalism, where it is proportional to the mesoscopic conductance and reads

\begin{equation}
\mathcal{T}=\textrm{Tr}\big\{\Gamma_LG^{r}\Gamma_RG^{a}\big\} \, .
\label{eq:5}
\end{equation}

Here $G^{r}$ and $G^{a}$ are the retarded and advanced dressed Green's functions at the angle disorder barrier. $\Gamma_{\{L,R\}}=i\big[\Sigma^r_{\{L,R\}}-\Sigma^a_{\{L,R\}}\big]$ are the coupling functions of the barrier to the pristine leads. These in turn depend on the self-energies of the leads $\Sigma_{\{L,R\}}$, i.e. the cumulative corrections to the bare Green's function at the barrier due to its coupling with them. To obtain these quantities, we develop a decimation technique based on Refs. \cite{cea2019twists,guinea1983effective} (see also Ref.~\cite{SM}). From the Green's function, the DOS also follows,

\begin{equation}
\textrm{DOS}=\frac{1}{\pi}\textrm{Tr}\big\{\textrm{Im}\{G(\omega)\}\big\} \, .
\label{eq:6}
\end{equation}

\begin{figure*}
    \centering
    {\includegraphics[scale=0.85]{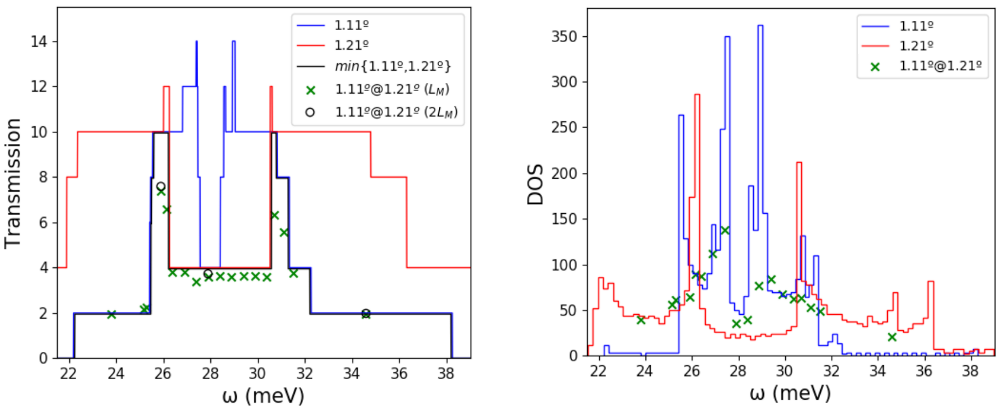}}%
    \caption{(Left) Transmission through an angle disorder barrier. Continuous lines are the bulk transmissions through pristine leads 1.11$^{\circ}$ and 1.21$^{\circ}$. Only carriers that can propagate on both angle domains may cross the angle disorder barrier, thus the transmission through the barrier 1.11$^{\circ}$@1.21$^{\circ}$ is bounded by $min\{\mathcal{T}_{1.11^{\circ}},\mathcal{T}_{1.21^{\circ}}$\}. Crosses (circles) correspond to transmission through a barrier of length one (two) moir\'e periods ($L_M)$. The transmission through the barrier is high all across the flat band, specially for carriers near $\Gamma$, $K$ and $K^{\prime}$. (Right) Local DOS at the disorder barrier 1.11$^{\circ}$@1.21$^{\circ}$ compared to the bulk DOS of its pristine leads. The van-Hove peaks are damped or erased at the barrier.}
    \label{fig:3}
\end{figure*}

{\it Results.} Figure \ref{fig:2} shows the band structures of TBG nanotube (56,2)@(-56-2), ($\theta=3.48^{\circ})$ scaled to simulate 1.11$^{\circ}$ and 1.21$^{\circ}$. The Brillouin zone of a TBG nanotube results from a folding of the Brillouin zone of TBG. In the case of (56,2)@(-56-2), the unit cell is four times larger than that of its associated TBG and this results in eight flat bands, not counting spin or valley degeneracy. More bands allow us to extract richer information from the transmission calculation. The bandwidth of the flat bands for the 1.11$^{\circ}$ system is approximately 16 meV, versus 44 meV for 1.21$^{\circ}$. The DOS and transmissions through the bulk of the pristine leads can be extracted directly from the bands; the number of transmission channels at a given energy is just number of band crossings at that energy and the DOS can be obtained with an histogram of the eigenvalues. The transmissions through the bulk of the pristine leads are plotted in Figure \ref{fig:3} with continuous lines. For the disorder barrier, the bands cannot be defined and using the decimation technique becomes a must. Figure \ref{fig:3} also shows the transmission through the angle barrier 1.11$^{\circ}$@1.21$^{\circ}$, for a barrier width of one or two moir\'e periods. Focusing first on the data points near 35 meV, we see that their transmission is virtually complete. These are normal incidence channels, but their perfect transmission is surprising since they live near the $\Gamma$ point (see also Figure \ref{fig:2}) and hence cannot experience Klein tunneling \cite{katsnelson2006}. Across the $\mathcal{T}=4$ plateau of 1.21$^{\circ}$, the barrier 1.11$^{\circ}$@1.21$^{\circ}$ allows transmissions in excess of 90\%. These are four Dirac-like channels, two normal and two oblique. Hence, we can infer that these oblique channels near $K$ and $K^{\prime}$ have transmissions over 80\% . Finally we observe that in the narrow window near 26 meV, where both pristine systems have ten transmission channels (two normal plus eight oblique), with all $k_\parallel$ values allowed by the periodic boundary condition, the transmission through the barrier still reaches 75\%. The new channels have noticeable electron-hole asymmetry, the transmission at the hole-like $\mathcal{T}=10$ window near 26 meV is $10\%$ higher than the at the electron-like window near 31 meV. Other angle combinations confirm this asymmetry in oblique channel transmission. Figure  ~\ref{fig:3} displays the bulk densities of states of pristine leads 1.11$^{\circ}$ and 1.21$^{\circ}$, together with the local density of states at the angle barrier. The pristine leads present several van-Hove peaks corresponding to the various flat bands seen in Figure \ref{fig:2}. At the angle disorder barrier, some van-Hove peaks are erased while the rest suffer pronounced damping.

\begin{figure}
    \centering
    {\includegraphics[scale=0.52]{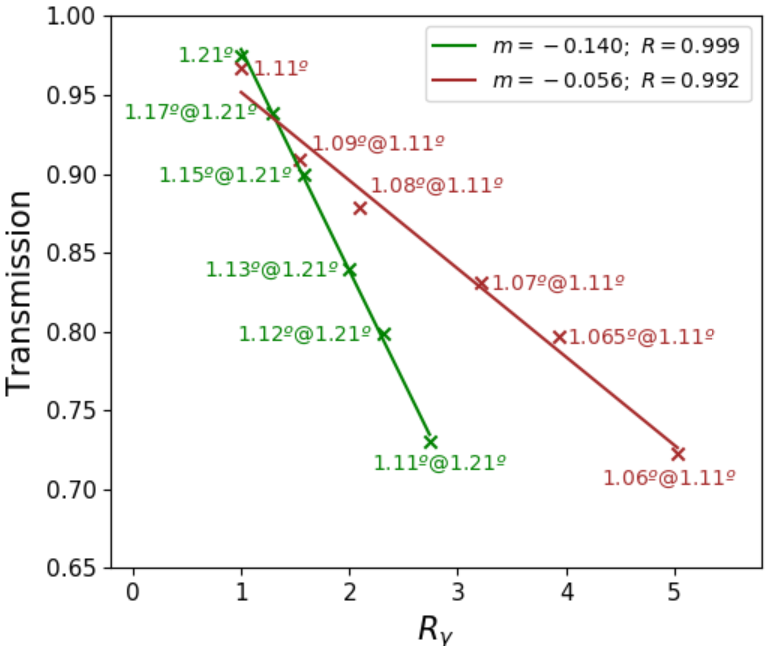}}
    \caption{Normalized transmission through several angle disorder barriers versus the bandwidth ratio of the connected angles. The values in each series are computed at the same point in energy, within a $\mathcal{T}=10$ window. The transmission decreases linearly with the bandwidth ratio.}
    \label{fig:5}
\end{figure}

The simulation of different angle combinations provides insight into the factors that determine the transmission. For angle barriers 1.32$^{\circ}$@1.21$^{\circ}$ and 1.06$^{\circ}$@1.11$^{\circ}$, figures analogous to Figure \ref{fig:3} can be found in Ref.~\cite{SM}.
In view of these, we conjecture that the bandwidth ratio of the connecting angles is the main quantity governing the transmission through a disorder barrier. The proximity to the magic angle and electron-hole asymmetry play secondary, albeit significant, roles. To verify this picture, we perform two series of simulations, shown in Figure \ref{fig:5}. The idea is to start with a barrier (e.g. 1.11$^{\circ}$@1.21$^{\circ}$) and reduce the angle difference in small steps, i.e. 1.12$^{\circ}$@1.21$^{\circ}$, 1.13$^{\circ}$@1.21$^{\circ}$, etc., each time computing the transmission within the $\mathcal{T}=10$ window at the same point in energy. The results are compelling, the transmission grows linearly as the bandwidth ratio diminishes.

{\it Discussion.} The results presented here for twisted bilayer graphene show that the transmission between regions with different twist angles is high, specially for incidence normal to the barrier between the regions. On the other hand, angle inhomogeneity significantly modifies the local density of states, broadening and suppressing features such as van Hove singularities (see also Ref.~\cite{wilson2020}). An interesting aspect is the complete transmission of normal incidence carriers, which occurs even near $\Gamma$ and contrasts with Ref. \cite{padhi2020}. This behaviour is very robust, normal incidence channels have virtually complete transmission ($\mathcal{T}\gtrsim95\%$) even for the most inhomogeneous barrier that we have studied, $1.11^{\circ}$@$3^{\circ}$, whereas the transmission of the oblique Dirac channels falls to $\mathcal{T}\lesssim5\%$ already for $1.11^{\circ}$@$1.9^{\circ}$. A straightforward description of this phenomenon in terms of Klein tunneling cannot be carried out, although it suggests that the orbital characters of the ingoing and outgoing waves are orthogonal. It is noteworthy that there is no reduction in transmission when doubling the width of the angle disorder barrier. This suggests that the transmission through angle disorder barriers could decay sub-exponentially with the barrier width, differing from previous predictions \cite{padhi2020}.

The transmissions through a barrier between regions twisted by different angles, and with ten channels is inversely proportional to the ratio between the bandwidths in the two electrodes. Future research could explore if this relation between bandwidth and transmission persists beyond our framework of angle disorder, and clarify if disorder in TBG samples can indeed be described as `bandwidth disorder', encompassing the effects of angle disorder, heterostrain and electron-electron interactions on the bandwidth. An interesting feature of Figure \ref{fig:5} is the different slope of each series. It is worth noting that the series starting with 1.11$^{\circ}$@1.21$^{\circ}$ was taken at the hole side of the spectrum, while the other series is from the electron side. However, there is a deeper reason for the slope mismatch: the series starting with  1.06$^{\circ}$@1.11$^{\circ}$ is closer to the magic angle. Ultimately, it is the matching of wavefunctions what determines the transmission through the barrier. As the  eigenfunctions change slowly near the magic angle, the transmission becomes less correlated with the bandwidth ratio of the angles and hence the smaller slope of the series. A toy model with single wall nanotubes reproduces all main results: in a single-wall nanotube with markedly different hoppings (hence bandwidths) in each lead, the transmission through the barrier is very notable, despite the disappearance of van-Hove peaks, see Ref. \cite{SM}. We have explored angle disorder within a single-particle picture. Hence a  shortcoming of the model is the absence of electron-electron interactions, thought to be of crucial importance for the outstanding transport phenomena seen in TBG. Moreover, our angle disorder is circumscribed to one or two moir\'e unit cells connected to semi-infinite pristine leads, whereas real systems have angle domains with many shapes and sizes. Also, due to computational constraints, the disorder modelled here is one to two orders of magnitude sharper than those seen in experiments, which have typical gradients of order $0.01^{\circ}-0.05^{\circ}/\mu m$. 

In conclusion, we have studied transport in TBG in the presence of angle disorder. To access the magic angle regime, we have done a scaling approximation. We have computed the Landauer-B\"uttiker transmission through an angle disorder barrier connected to pristine leads, employing a decimation technique. The van-Hove peaks of the local DOS at the barrier are very sensitive and can disappear with angle disorder of $\sim0.1^{\circ}$. In contrast, the transmission through the barrier is remarkably high all across the flat bands, and very close to one for normal incidence. Within high-$\mathcal{T}$ windows, the transmission through an angle disorder barrier is, to first order, inversely proportional to the bandwidth ratio of the angles it connects. This work lays the foundations for future studies on the effects of disorder on transport in TBG. A crucial step will be to understand the interplay between electron interactions and disorder. Angle disorder is omnipresent in moir\'e systems and is likely to have immediate as well as subtle consequences on their phase diagrams, only future research will tell how far-reaching are its ramifications in this new field of condensed matter physics.

{\it Acknowledgements}
This work was supported by funding from the European Commission, within the Graphene Flagship, Core 3, grant number 881603 and from grants NMAT2D (Comunidad de Madrid, Spain), SprQuMat and SEV-2016-0686 (Ministerio de Ciencia e Innovación, Spain).

\bibliography{AngleDisorder}
\clearpage
\onecolumngrid


\section*{Supplementary material}
\setcounter{equation}{0}
\setcounter{figure}{0}
\setcounter{table}{0}
\makeatletter
\renewcommand{\theequation}{S\arabic{equation}}
\renewcommand{\thefigure}{S\arabic{figure}}

\section*{S1. Additional results}
\subsection{Toy model of disorder}

\begin{figure}[h]
    \centering
    {\includegraphics[scale=0.650]{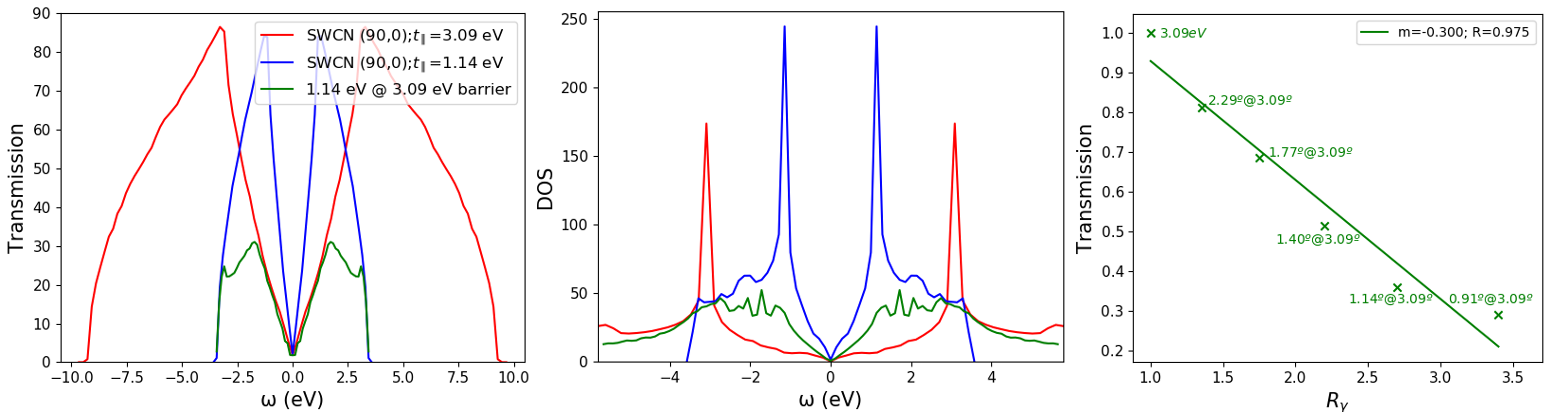}}
    \caption{Wide band -- narrow band toy model. (Left) Transmission through zig-zag nanotubes (90,0) with $t_{\parallel}=3.09$ eV and $t_{\parallel}=1.14$ eV, compared to the transmission through a hopping-disorder barrier that connects both. The width of the barrier is 10 unit cells. There is notable transmission when the narrow band activates. (Centre) Bulk DOS of the pristine nanotubes and local DOS at the hopping-disorder barrier. Disorder erases the van-Hove peaks. (Right) Normalized peak transmission at several 10-cell hopping-disorder barriers against the bandwidth ratio of their pristine leads, full data sets in Figure \ref{fig:S4}.}
    \label{fig:6}
\end{figure}

When TBG carriers pass from a region with 1.11$^{\circ}$ to another with 1.21$^{\circ}$ through an angle disorder barrier, they see only a slight difference in moir\'e period. In comparison, the bandwidth augments dramatically by a factor $R_{\gamma}=2.7$. Therefore, we can try to capture the leading physical effect with a toy model of a wide band system connected to another with a narrower band. To that end, we consider a single-wall zig-zag nanotube with chiral vector (90,0), one lead with nearest neighbour hopping $t_{\parallel}=3.09$ eV and the other with $t_{\parallel}=1.14$ eV, connected through a hopping disorder barrier with a width of ten unit cells. Figure \ref{fig:6} shows the resulting transmission and DOS. Despite the sudden bandwidth reduction, there is considerable transmission through the barrier when the narrow band activates. Just like in TBG with angle disorder, van-Hove peaks are largely erased in the presence of disorder, but carrier transmission remains remarkably high. Furthermore, the right panel shows the normalized peak transmission through various hopping disorder barriers versus the bandwidth ratio of their pristine leads. The toy model reproduces well the linear relation between transmission and bandwidth ratio of the leads.

\begin{figure}[h]
\vspace*{+0.4cm}
\hspace*{-0.6cm} 
    \centering
    {{\includegraphics[width=6.5cm]{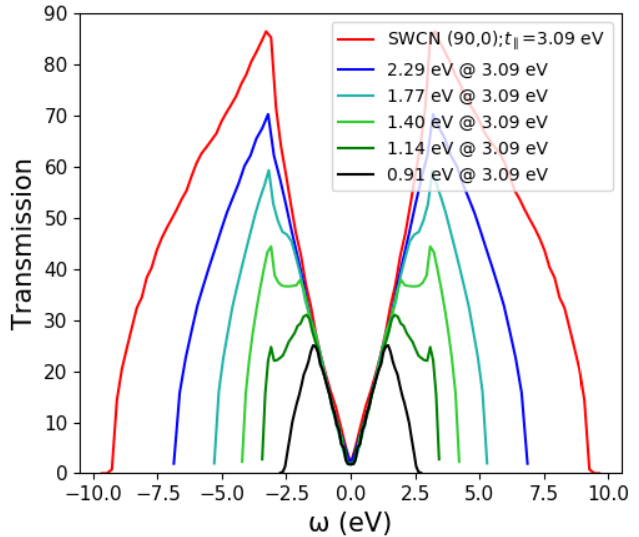} }}%
    \qquad
    {{\includegraphics[width=6.5cm]{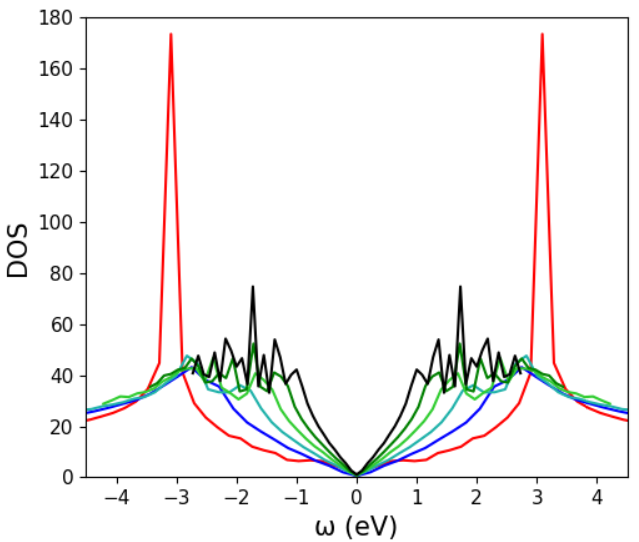} }}%
    \caption{Graphene as a toy model of disorder. Transmissions (left) and DOS (right) through hopping disordered barriers $A@B$ that connect pairs of single wall carbon nanotubes with different hoppings $A$ and $B$.}%
    \label{fig:S4}
\end{figure}

Figure \ref{fig:S4} shows the DOS and transmissions through several toy model barriers, including the one in Figure \ref{fig:6}, 1.14 eV @ 3.09 eV. In the DOS, the original van-Hove peaks of the systems that connect are eliminated at the barriers, although smaller resonances appear and grow in number as disorder increases. In the transmission, we can distinguish three regimes: for low disorder, there is a single maximum in the disordered transmission which comes from the broad band system (e.g. 2.29 eV @ 3.09 eV and 1.77 eV @ 3.09 eV); for medium disorder, the disordered transmission has two maxima, inherited from both the broad and narrow band systems (e.g. 1.40 eV @ 3.09 eV and 1.14 eV @ 3.09 eV); for large disorder, the peak due to the broad band system is lost, and only the one coming from the narrow system remains (0.91 eV @ 3.09 eV). The maxima of each transmission curve provide the data for the right panel of Figure \ref{fig:6}. There we see that the relation between transmission and bandwidth ratio is approximately linear, but not strictly so. This is because the existence of several regimes results in two different slopes, one while the maximum comes from the broad band system and other when it starts to originate from the narrow band system.

\subsection{TBG with angle disorder: Klein tunneling}

Normal incidence carriers transmit without back-scattering ($\mathcal{T}\approx1$) through the angle barrier 1.11$^{\circ}$@1.21$^{\circ}$. One naturally wonders how robust this effect is. To answer this, we have looked at more inhomogeneous barriers, up to 1.11$^{\circ}$@3.00$^{\circ}$. For each barrier in the table below, we compute the transmission of the two normal incidence channels near $\Gamma$ (i.e. we recalculate the data point near 35 meV in Figure \textcolor{red}{3} in the main text) and the transmission of the Dirac mixture (two normal plus two oblique channels with $k_{\parallel}=\sqrt{3}k_{\perp}$) near the charge neutrality point. With these, one can infer the transmission of the oblique incidence channels. There is a stark difference between normal and oblique incidence channels. Normal channels are almost impervious to disorder and reach $\mathcal{T}\approx1$ for all barriers. In contrast, the transmission of oblique incidence channels is very low already for 1.11$^{\circ}$@1.90$^{\circ}$. It is worth noting that for abrupt barriers like this one, we cannot justify the approximation that neglects changes in the moir\'e period and which allowed us to model angle disorder as intralayer hopping disorder. Hence, for abrupt barriers we need to bear in mind that the intralayer hopping disorder we introduce can no longer faithfully mimic angle disorder. In any case, the chiral nature of carriers in graphene, which  enables Klein tunneling through $n-p$ \cite{cheianov2006selective} and $n-p-n$~\cite{katsnelson2006} potential barriers, could be forbidding back-scattering of normal incidence carriers in TBG too. An immediate consequence is that transport across twist angle domains leads to the collimation of carrier momenta around normal incidence.

\begin{center}
\vspace{+0.5cm}
\begin{tabular}{|p{3cm}||p{3cm}|p{3cm}|p{3cm}|p{3.5cm}|}
 \hline
 \multicolumn{5}{|c|}{Transmision with angle disorder} \\
 \hline
 Barrier $A@B$ & Mini-gaps of $B$ &Normal-channel $\mathcal{T}$ & Dirac mixture $\mathcal{T}$ & Oblique $k_{\parallel}=\sqrt{3}k_{\perp}$ $\mathcal{T}$\\
 \hline
 1.11$^{\circ}$@1.21$^{\circ}$   & Two    &0.99&   0.90&   0.80\\
 1.11$^{\circ}$@1.41$^{\circ}$   & Two    &0.98&      0.66&   0.35\\
 1.11$^{\circ}$@1.60$^{\circ}$   & Two    &0.97&      0.56&   0.16\\
 1.11$^{\circ}$@1.90$^{\circ}$   & One    &0.96&   0.50&   0.03\\
 1.11$^{\circ}$@3.00$^{\circ}$  & None   &0.95&   0.46&   $<$0.01\\
 \hline
\end{tabular}
\end{center}
\begin{figure}[h]

\hspace*{-0.3cm}
    \centering
     {\includegraphics[width=6.5cm]{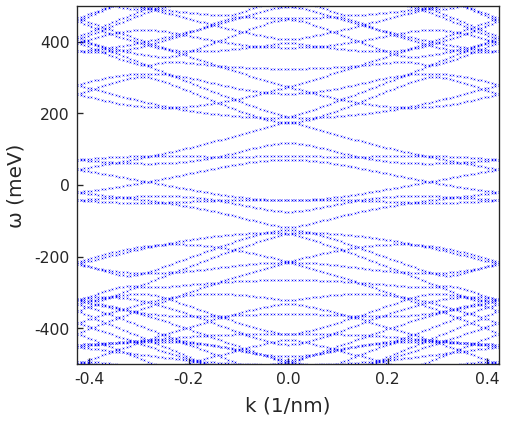}}%
     {\includegraphics[width=6.5cm]{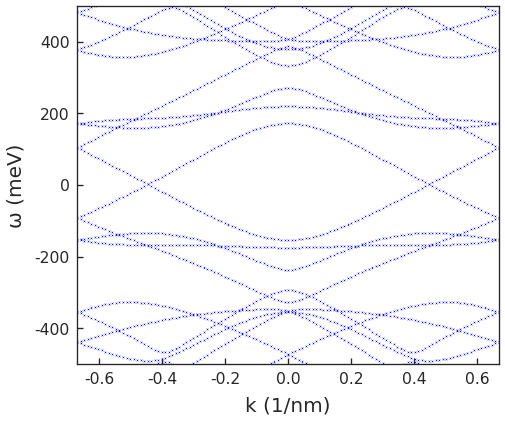}}%
    \caption{(Left) $\theta=1.90^{\circ}$ bands. (Right) $\theta=3.00^{\circ}$ bands.}
    \label{fig:S1}
\end{figure}
\newpage

\subsection{TBG with angle disorder: Other barriers}
\begin{figure}[h]
    \centering
    {{\includegraphics[width=6.5cm]{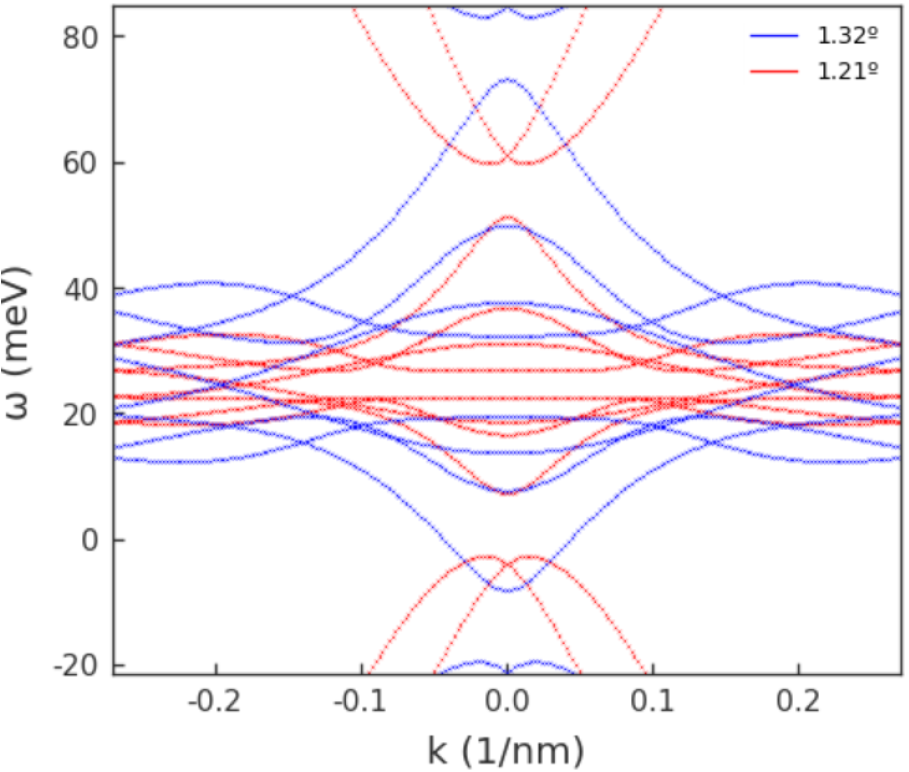} }}%
    \qquad
    {{\includegraphics[width=6.5cm]{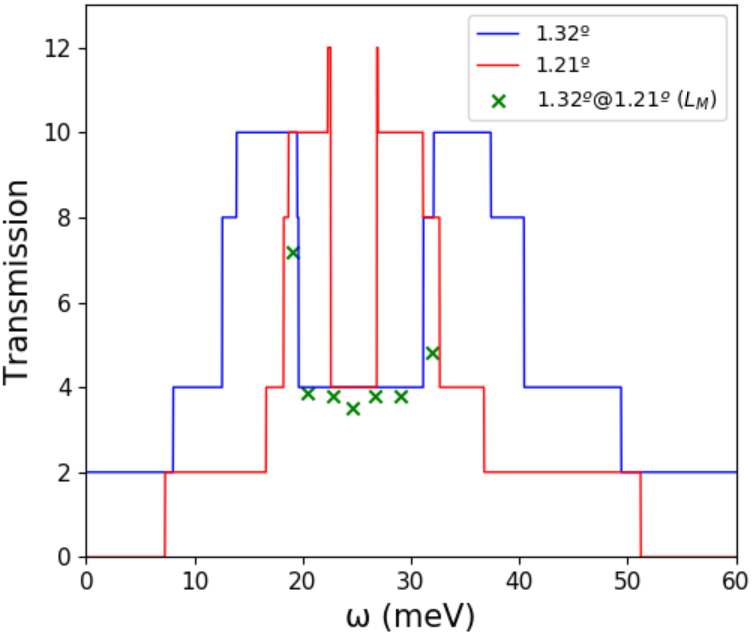} }}%
    \caption{(Left) Band structures of TBG nanotubes (56,2)@(-56,-2) ($\theta=3.48^{\circ}$) scaled to simulate 1.32$^{\circ}$ and 1.21$^{\circ}$. (Right) Transmission through the angle disorder barrier 1.32$^{\circ}$@1.21$^{\circ}$, compared to the bulk transmission through its pristine leads.}%
    \label{fig:S2}
\end{figure}

\begin{figure}[h]
    \centering
    {{\includegraphics[width=6.4cm]{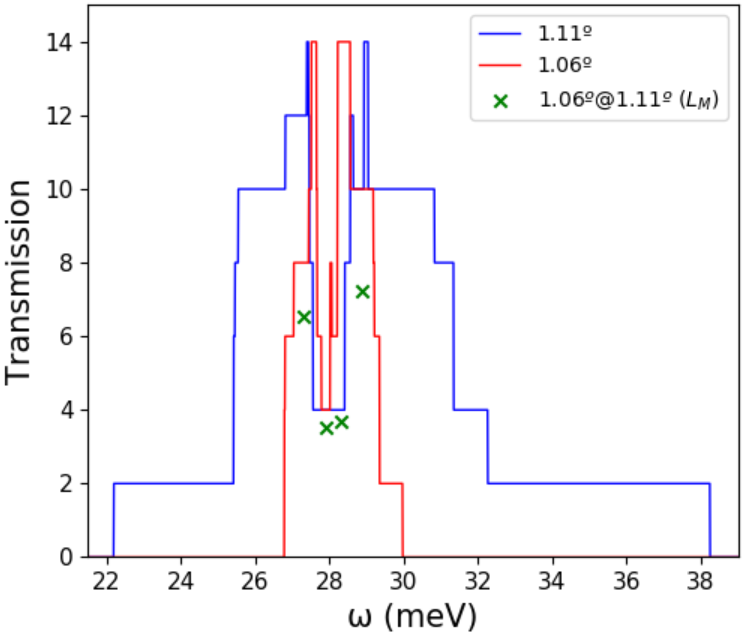} }}%
    \qquad
    {{\includegraphics[width=6.35cm]{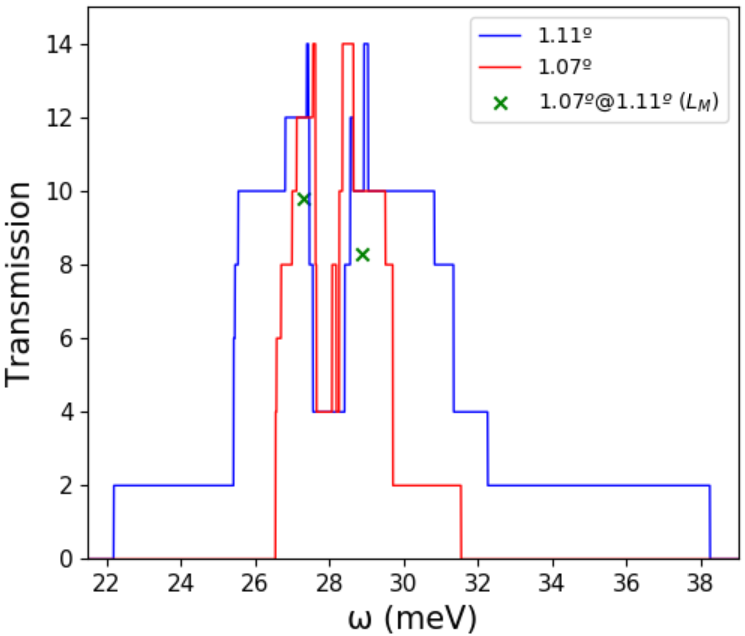} }}%
    \caption{(Left) Transmission through the angle disorder barrier 1.06$^{\circ}$@1.11$^{\circ}$, compared to those of its pristine leads. (Right) Same for 1.07$^{\circ}$@1.11$^{\circ}$.}%
    \label{fig:S3}
\end{figure}

In the main text we focus on the barrier 1.11$^{\circ}$@1.21$^{\circ}$ \footnote{We have rounded 1.114$^{\circ}$ to 1.11$^{\circ}$ and 1.206$^{\circ}$ to 1.21$^{\circ}$ for clarity.}. The transmission and densities of states are obtained with the decimation technique outlined in section S4. Figures \ref{fig:S2} and \ref{fig:S3} present the results of analogous computations for 1.32$^{\circ}$@1.21$^{\circ}$, 1.06$^{\circ}$@1.11$^{\circ}$ and 1.07$^{\circ}$@1.11$^{\circ}$ and complement the discussion of the main text about the factors that influence transmission. The similar values obtained for barriers 1.11$^{\circ}$@1.21$^{\circ}$, 1.32$^{\circ}$@1.21$^{\circ}$ and 1.06$^{\circ}$@1.11$^{\circ}$, suggest that the bandwidth ratio of the angles that the barrier connects is the most important factor determining transmission through it. Other relevant considerations are the proximity to the magic angle and the existence of electron-hole asymmetry in favour of the holes. Together, these factors form a nuanced picture, but the major effect of the bandwidth ratio can be distilled from the rest by performing simulations in which larger angle is fixed and the smaller increases until they match, see main text. During the procedure, the bandwidth ratio changes fast while the other factors do not; the proximity to the magic angle stays similar and the transmission is computed at the exact same energy, avoiding $e^{-}-h^{+}$ asymmetry. The transmission is found to be inversely proportional to the bandwidth ratio of the connecting angles.

The $e^{-}-h^{+}$ asymmetry appears in the transmission of oblique incidence channels, within high-$\mathcal{T}$ windows. Studying the asymmetry systematically is difficult because it usually makes such windows reach different values on the electron and hole sides, until the coupled angles become sufficiently similar and transmissions are too close to 1 for a meaningful comparison. Nevertheless, comparing windows with different $\mathcal{T}$ is still informative, bearing in mind that reaching higher $\mathcal{T}$ implies adding more oblique incidence channels, which are in principle less capable of crossing the barrier. For example, the transmission is $60\%$ within the $\mathcal{T}=8$, $e^{-}$ window of 1.32$^{\circ}$@1.21$^{\circ}$ (near $\omega=33$ meV in Figure \ref{fig:S2}), while it reaches $72\%$ within the $\mathcal{T}=10$, $h^{+}$ window near 20 meV. Because of the two extra channels, the $\sim12\%$ edge of the holes has more merit. In contrast, the barrier 1.06$^{\circ}$@1.11$^{\circ}$ has $82\%$ transmission at a $\mathcal{T}=8$, $h^{+}$ window and $72\%$ at a $\mathcal{T}=10$, $e^{-}$ window. For 1.07$^{\circ}$@1.11$^{\circ}$ one finds $\sim82\%$ for both $\mathcal{T}=12$, $h^{+}$ and $\mathcal{T}=10$, $e^{-}$ windows. Overall, these results suggest that there is less $e^{-}-h^{+}$ asymmetry near the magic angle.

\subsection{Integrated charge map of TBG}

\begin{figure}[h]
    \centering
    {{\includegraphics[width=9cm]{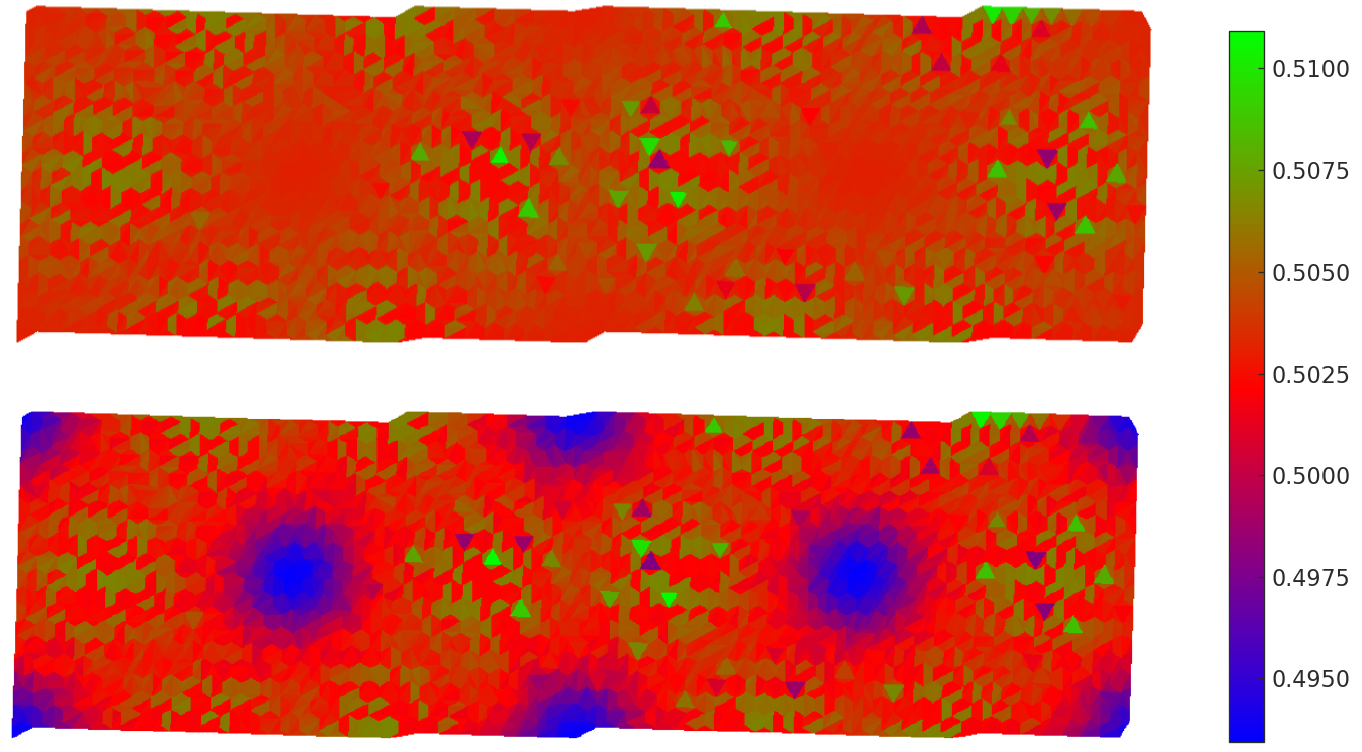} }}%
    \caption{(Bottom) 1.11$^{\circ}$ TBG charge map integrated up to the gap between hole-like remote bands and flat bands. AA regions have a charge deficit of $\sim2\%$ relative to AB. (Top) Integrating further, up to the charge neutrality point, the charge balances between AA and AB stacking regions (relative deviations are $\lesssim0.5\%$).
    \label{fig:S5}}
\end{figure}

Finally, we looked at integrated charge maps. In Figure \ref{fig:S5} we plot the integrated charge map in the unit cell of a pristine system with $\theta=1.11^{\circ}$, obtained from exact diagonalization. The bottom map is integrated up to the gap between the hole-like remote bands and the flat bands, and this results in AA regions having $\sim2\%$ less accumulated charge than AB regions. The top map shows that this deficit is compensated when we include the hole-like flat bands, i.e. when we integrate all bands up to the charge neutrality point. At the angle disorder barrier, the charge map (not shown) approximately corresponds to a smooth interpolation of the maps on its pristine leads.

\section*{S2. Real lattice}

\begin{figure}[h]
\hspace*{-0.3cm}
    \centering
    \begin{subfigure}
     {\includegraphics[width=7.5cm]{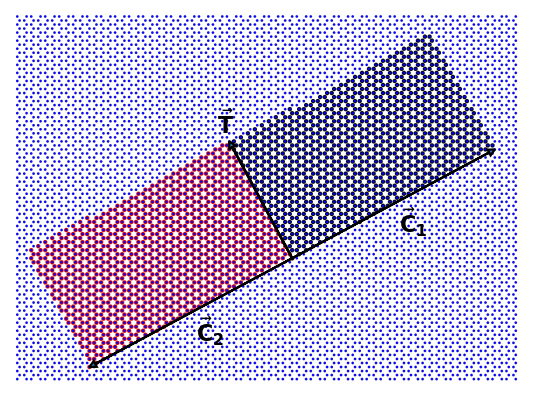}}%
    \end{subfigure}
    \begin{subfigure}
     {\includegraphics[width=4cm]{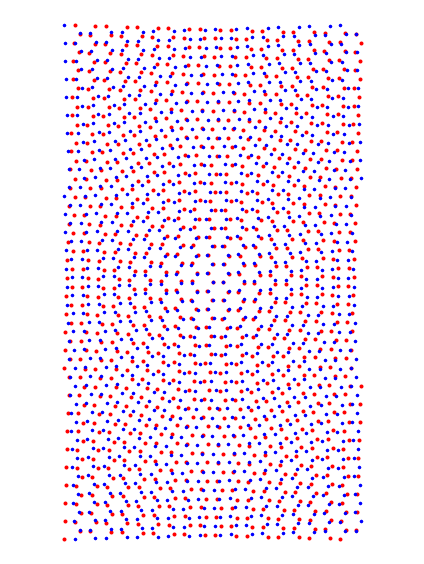}}%
    \end{subfigure}
    \caption{Construction of a twisted bilayer graphene nanotube (28,1)@(-28,-1), a system with a twist angle of $3.48^{\circ}$. (Left) Graphene honeycomb lattice with highlighted units cells of nanotubes with chiral vectors $\vec{C_1}=(n,m)=(28,1)$ (dark blue) and $\vec{C_2}=(-28,-1)$ (red), the chiral angle is $\theta_C\approx1.74^{\circ}$ in both cases. Also drawn is their translational vector $\vec{T}.$ (Right) Unit cell of a twisted bilayer graphene nanotube (28,1)@(-28-1), obtained from rotation of the cells in the left panel by $120^{\circ}-\theta_C$ followed by reflection of the (-28,-1) cell with respect to the x axis. The resulting twist angle is $\theta=2\cdot\theta_C$.}
    \label{fig:S6}
\end{figure}

The geometry of a carbon nanotube is uniquely defined by its chiral vector $\vec{C}=n\vec{a}_1+m\vec{a}_2$, which specifies a possible wrapping of the graphene lattice into tubular form. The chiral vector determines the diameter of the nanotube and the number of atoms $N_C$ in its translational unit cell. The chiral angle $\theta_C$ is defined with respect to the basis vector $\vec{a}_1=(\sqrt{3}/2,1/2)a$, which forms $30^{\circ}$ with the horizontal and is the chiral vector of zig-zag nanotubes. Figure \ref{fig:S6} depicts a way to construct a `twisted bilayer graphene nanotube', a system of two concentric nanotubes with opposite chiral vectors that we use to mimic TBG. It should be noted that Figure \ref{fig:S6}(R) is a planar representation of the system's unit cell, and that the TBG nanotube results from imposing periodic boundary conditions in the vertical direction.

We include below the expressions of the two most relevant quantities for this research that follow from the geometry. These are the twist angle $\theta$ and the number of atoms N in the translational unit cell of a (n,m)@(-n,-m) TBG nanotube, as a function of $n$ and $m$ \cite{charlier2007electronic},

\begin{equation}
\theta= 2\cdot arccos\Big(\frac{2n+m}{2\sqrt{n^2+nm+m^2}}\Big)\, ,
\label{eq:S1}
\end{equation}

\begin{equation}
N=\frac{8(n^2+nm+m^2)}{gcd(2m+n,2n+m)}\, .
\label{eq:S2}
\end{equation}

It is worth mentioning that the system built in Figure \ref{fig:S6} is one of the smallest possible on which it is feasible to apply a scaling procedure plus a decimation technique to calculate the transmission with disorder near the magic angle ($\theta \sim 1^{\circ}$). In particular, (28,1)@(-28,-1) requires a scaling factor $\lambda \sim 3.5$ and finite size effects appear beyond $\lambda \sim 4.5$. However, scalings up to $\lambda \sim 7$ are still useful for exact diagonalization methods that use many unit cells.  For more details see the sections on Scaling and Decimation below. 

\section*{S3. Reciprocal lattice}

\begin{figure}[h]
\hspace*{-0.2cm}
    \centering
    \begin{subfigure}
     {\includegraphics[width=5.5cm]{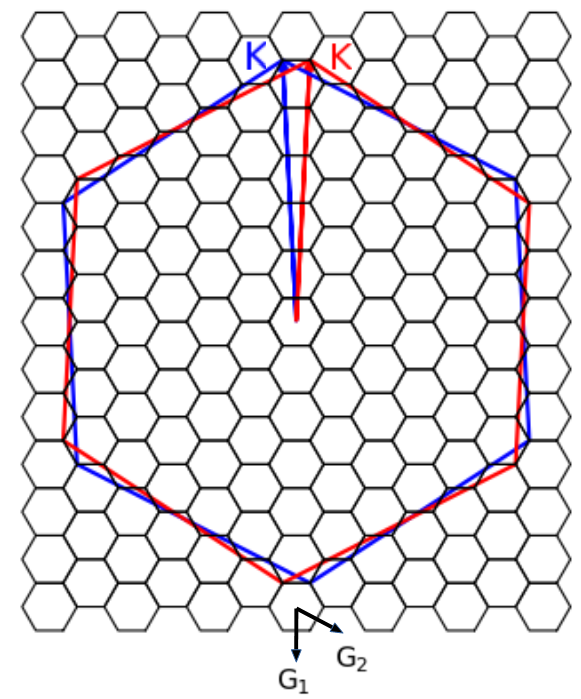}}%
    \end{subfigure}
    \begin{subfigure}
     {\includegraphics[width=5.5cm]{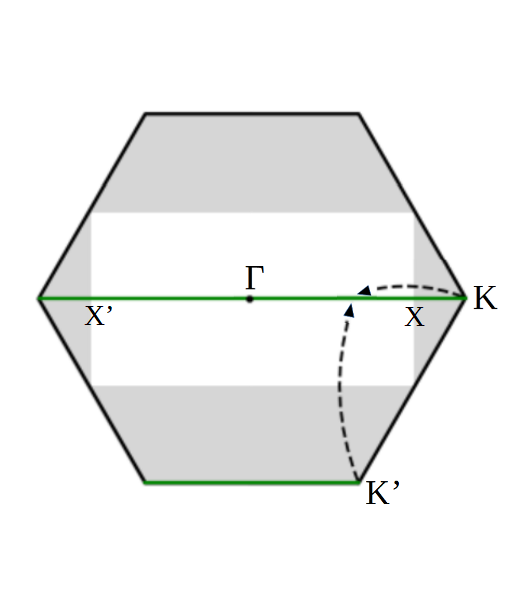}}%
    \end{subfigure}
    \caption{Construction of the 1D Brillouin zone of a TBG nanotube (n,1)@(-n,-1). (Left) Construction of the mBZ of TBG, done with a large $\theta$=$6.01^{\circ}$ for clarity, the blue and red hexagons are the Brillouin zones of each graphene layer. The K point displacement determines the size of the mBZ. When the twist angle is commensurate, one can do a tiling of the mBZ. (Right) mBZ of TBG, BZ of an unrolled TBG nanotube (n,1)@(-n,-1) (white) and allowed $k_y$ (green lines). To obtain the bands of the TBG nanotube, one folds the mBZ onto the white rectangle and superimposes the TBG bands along the green lines.}
    \label{fig:S7}
\end{figure}

Because a TBG nanotube is a one-dimensional system, its Brillouin zone is one-dimensional too. Indeed, it is the same Brillouin zone as the one of its constituent single-wall nanotubes, namely the segment $X^{\prime}-\Gamma-X$ in Figure \ref{fig:S7}(R), of length $2\pi/|T|$. More insightful is the observation that the bands of a TBG nanotube can be obtained from a folding \cite{charlier2007electronic} of the bands of planar TBG, bearing in mind momentum quantization. This is what we aim to show in this section.

Consider the unit cell given by the dark blue lattice in Figure \ref{fig:S6}(L). Its reciprocal basis vectors follow from the well known conditions $\vec{a_i}\cdot \vec{b_j}=2\pi\delta_{ij}$, with $\vec{a_1}=\vec{C_1}$ and $\vec{a_2}=\vec{T}$. After rotation (plus reflection in the case of the (-n,-m) lattice), the reciprocal unit cell of such lattice is in both cases (before actually rolling them into nanotubes) the white rectangle in Figure \ref{fig:S7}(L), with length $2\pi/|T|$ and width $2\pi/|C|$. For TBG nanotubes with chiral angle $\theta$, such rectangle is inscribed in the mini Brillouin zone (mBZ) of planar TBG with twist angle $\theta$.

Let the horizontal axis be the direction of propagation, rolling the lattice to obtain the true nanotube unit cell imposes periodic boundary conditions which quantize the momentum in the vertical direction. Therefore, starting from the mBZ of planar TBG with twist angle $\theta$ in Figure \ref{fig:S7}(R), the 1D Brillouin zone and band structure of the TBG nanotube with angle $\theta$ can be obtained by folding the shaded corners of the mBZ onto the white rectangle and superimposing the TBG energy bands along the green lines, which specify the $k_y$ momenta allowed by the periodic boundary condition. It is worth noting that the folding places the $K$ and $K^{\prime}$ points of the mBZ at two thirds of $\Gamma-X$. It is straightforward to prove that any TBG nanotube (n,1)@(-n,-1) will have allowed $k_y$ lines that pass through the center and sides of the mBZ of the corresponding planar TBG. Here we set $a=1$. Starting from the periodic boundary condition $\psi(\vec{r}+|C|\hat{u}_y)=e^{ik_y|C|}\psi(\vec{r})=\psi(\vec{r})$, one obtains the spacing between permitted $k_y$, $\Delta{k_y}=2\pi/|C|=2\pi/\sqrt{n^2+m^2+mn}$. On the other hand, the K point relative displacement is $\frac{4\pi}{3}\sqrt{2-2cos(\theta)}$ and the apothem of the mBZ is therefore $\frac{2\pi}{\sqrt{3}}\sqrt{2-2cos(\theta)}$. Substituting the expression for $\theta$ in (1) in the previous formula and using trigonometric identities the apothem reduces to $2\pi\sqrt{m^2/(n^2+m^2+mn)}$, which for m=1 is equal to the $k_y$ spacing obtained before.

The band folding argument explains the connection between the bands of a TBG nanotube and planar TBG. For practical purposes, the bandstructure of a TBG nanotube can also be calculated by diagonalization of the Hamiltonian matrix of the unit cell in Figure \ref{fig:S6}(R). One should impose periodic boundary conditions from top to bottom (rolling the nanotube) and from left to right and across the diagonals (reflecting the fact that the unit cell repeats in the horizontal axis). Also, Bloch's theorem dictates that each term in the Hamiltonian should be multiplied by a phase factor $e^{ik_x\Delta x}$, where $k_x$ is the continuum momentum along the propagation direction and $\Delta x$ is the horizontal distance between the sites that the interact through that particular term in the Hamiltonian.

\section*{S4. Scaling}
As already noted by Bistritzer and MacDonald \cite{bistritzer2011moire} the bands of twisted bilayer graphene depend on a single parameter, which involves the ratio of interlayer to intralayer hopping and the twist angle. For small angles we can write:

\begin{equation}
\alpha = \frac{at_{\perp}}{2\hbar v_F \sin ( \theta / 2 )} \propto \frac{t_{\perp}}{t_{\parallel}\theta}\, .
\label{eq:S3}
\end{equation}

One can also understand this as a comparison between the time a carrier takes to traverse the moir\'e unit cell versus the average time between interlayer tunneling events. In virtue of (\ref{eq:S3}), it is possible to perform a scaling approximation \cite{gonzalez2017} that allows us to reproduce the spectrum at the magic angle $\theta\sim1^{\circ}$ using a larger angle $\theta^{\prime}$ with a reduction in $t_{\parallel}$. The lattice should be magnified for consistency (e.g. so that the moir\'e period of a $1^{\circ}$ system is $\approx13$ nm). More formally, scaling is possible because the Dirac equation that governs each layer can be described in different ways, and our approach is a description based on a lattice of `super-atoms' with renormalized intralayer hopping and also a renormalized lattice constant. Thus, the scaling transformations are,

\begin{equation}
t_{\parallel}\rightarrow\frac{1}{\lambda} t_{\parallel}
\label{eq:S4}
\end{equation}
\begin{equation}
a\rightarrow\lambda a 
\label{eq:S5}
\end{equation}
\begin{equation}
d\rightarrow\lambda d  
\label{eq:S6}
\end{equation}
with
\begin{equation}
\lambda=\frac{\sin(\frac{\theta^{\prime}}{2})}{\sin(\frac{\theta}{2})}\, .  
\label{eq:S7}
\end{equation}

\begin{figure}[h]
    \centering
    {{\includegraphics[width=7cm]{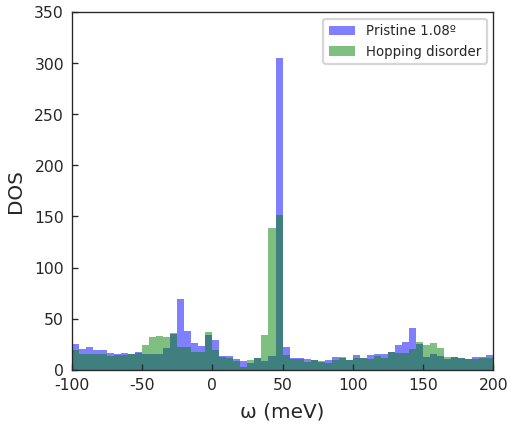} }}%
    \qquad
    {{\includegraphics[width=7cm]{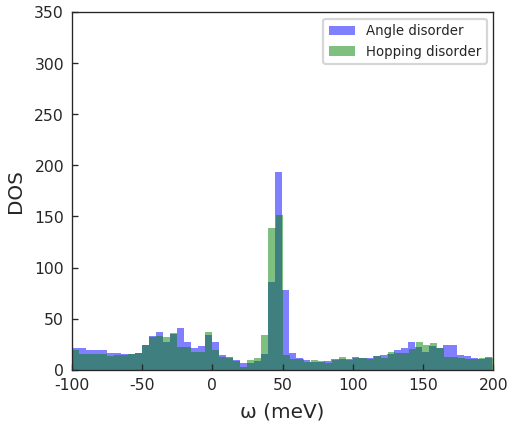} }}%
    \caption{(Left) DOS of a TBG system with uniform twist angle $\theta=1.08^{\circ}$ (blue) compared to that of a system with intralayer hopping disorder simulating $1.08^{\circ}$-$1.20^{\circ}$ twist angle disorder (green). (Right) Intralayer hopping disorder DOS (green) compared to true twist angle disorder (blue).}%
    \label{fig:S8}
\end{figure}

Furthermore, Eq. (\ref{eq:S3}) invites us to model twist angle disorder as intralayer hopping disorder, which is advantageous because it allows us to work with an undistorted lattice in the calculation of the transmission. Next we include some tests aimed at verifying this idea. Figure \ref{fig:S8}(L) shows the low energy density of states (DOS) of a pristine magic angle system ($\theta=1.08^{\circ}$) simulated with $\theta^{\prime}=5.40^{\circ}$ i.e. with a scaling factor $\lambda=5$, compared to the density of states of a system in which intralayer hopping disorder simulates twist angle disorder in the range $1.08^{\circ}$-$1.20^{\circ}$. Disorder damps and broadens the magic angle van-Hove peaks.
Next we compare intralayer hopping disorder to twist angle disorder. Since angle disorder deforms the lattice, we need to substitute the constant intralayer hopping in Eq. (\textcolor{red}{2}) in the main text with an expression that accounts for distortions: $\gamma_{ij}^{mm}=t_{\parallel}e^{-(r-a_{cc})/\lambda_\parallel}$, where $a_{cc}=0.142$ nm is the carbon-carbon distance and $\lambda_{\parallel}=0.042$ nm is a cutoff \cite{pereira2009tight}. Figure \ref{fig:S8}(R) shows the comparison between the system with intralayer hopping disorder and one with twist angle disorder. There is good qualitative agreement between both types of disorder. 

\begin{figure}[h]
    \centering
    {{\includegraphics[width=7cm]{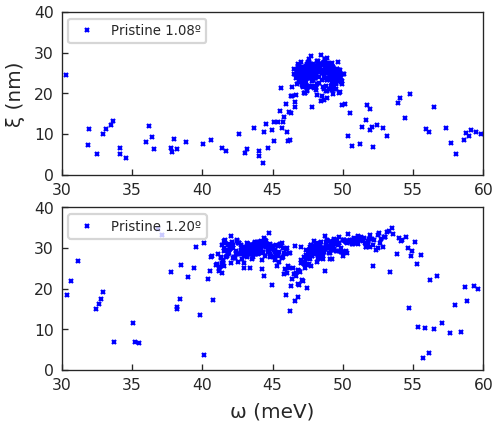} }}%
    \qquad
    {{\includegraphics[width=7cm]{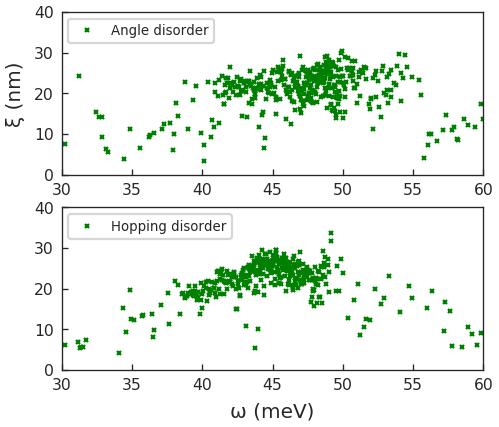} }}%
    \caption{(Left) Localization length of states in pristine $1.08^{\circ}$ and $1.20^{\circ}$ TBG systems. (Right) Localization length of TBG systems with twist angle disorder in the range $1.08^{\circ}$-$1.20^{\circ}$ and equivalent intralayer hopping disorder.}%
    \label{fig:S9}
\end{figure}

\begin{figure}[h]
    \centering
    {{\includegraphics[width=13cm]{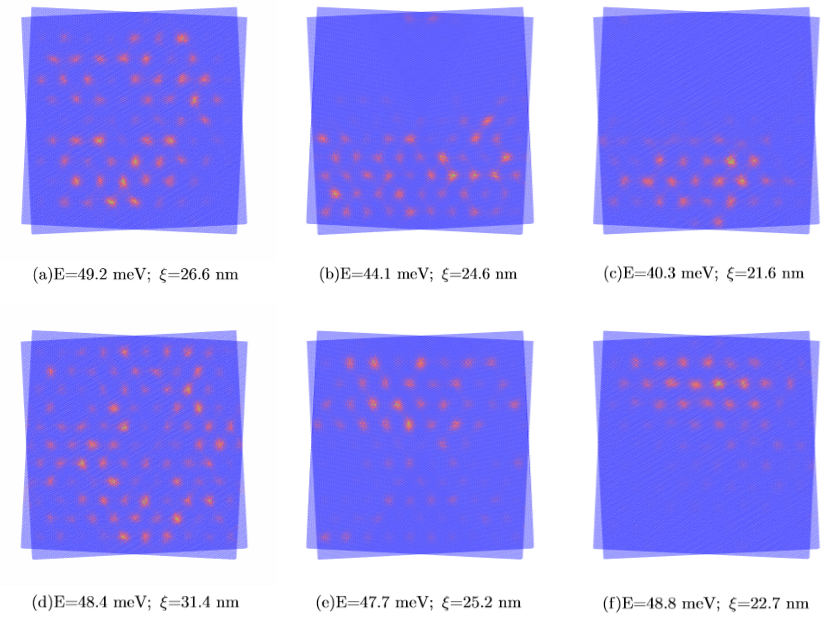} }}%
    \caption{(a,d) Charge maps of states in pristine $1.08^{\circ}$ and $1.20^{\circ}$ TBG respectively, with charge distributed across the sample in the AA stacking regions. (b,e) Maps of states in TBG with twist angle disorder in the range ($1.08^{\circ}$-$1.20^{\circ}$) showing charge localized in either angle domain. (c,f) Maps of states in TBG with intralayer hopping disorder simulating twist angle disorder, charge accumulates in a hopping domain.}%
    \label{fig:S10}
\end{figure}

 Disorder is introduced varying the twist angle (or the intralayer hopping) with an hyperbolic tangent function over the sample; the function changes over an extension large with respect to the moir\'e period. These and the following results (Figures \ref{fig:S8}-\ref{fig:S10}) were obtained by exact diagonalization of a TBG system with $\sim 44000$ lattice sites, which for these twist angles gives rise to approximately 80 moir\'es. Because of the ensuing distortion of the lattice in the system with twist angle disorder, it is simpler here to omit the periodic boundary conditions, however we checked that for this system size, their inclusion has a negligible effect. 

We use the Inverse Participation Ratio (IPR) and its associated measure, the localization length, as another measure of similitude between both types of disorder. The IPR of a state `i' is defined as the sum of the fourth powers of its eigenvector components \cite{kramer1993localization}:

\begin{equation}
IPR_i=\sum_{k}{|\phi_{i,k}|^4}\, ,
\label{eq:S8}
\end{equation}

from which the localization length follows:
\begin{equation}
\xi_i=\frac{1}{2}\sqrt{\frac{1}{IPR_i}}\, .
\label{eq:S9}
\end{equation}

In Figure \ref{fig:S9}(L) we present a comparison of the localization length of states in pristine TBG systems with $1.08^{\circ}$ and $1.20^{\circ}$. The latter shows two well differentiated clusters corresponding to the van Hove peaks, which in the former are merged into one, indicating the close proximity of $1.08^{\circ}$ to the magic angle in our model. Figure \ref{fig:S9}(R) shows the localization of states in systems with angle disorder in the range ($1.08^{\circ}$-$1.20^{\circ}$) and intralayer hopping disorder aiming to mimic that angle disorder. Compared to the pristine systems, the average localization decreases slightly and the two clusters merge. There is also a considerable broadening of the bandwidth in comparison with the $1.08^{\circ}$ system. Hopping disorder leads to a slight redshift of the flat bands compared to angle disorder and it also results in smaller localization lenght on average, hence in a sense it is a more constraining type of disorder.

Finally, we studied the real space charge localization maps of pristine, angle disordered and hopping disordered systems. Some characteristic states are plotted in Figure \ref{fig:S10}. Most of the states in a pristine systems tend to spread over the whole sample, with charge accumulating at the AA stacking regions. In contrast, systems with twist angle (or intralayer hopping) disorder lead to states in which charge accumulates preferably at an angle (or hopping) domain, which is consistent with their lower average localization length seen in Figure \ref{fig:S9}. 

Inspecting the charge maps of all states in the disordered systems, one sees that within certain energy intervals, a type of state prevails. For example, in the TBG with angle disorder, in the interval (41 meV, 45 meV) almost all states show charge accumulation in the lower half of the sample (which has larger twist angle $1.20^{\circ}$), whereas in the interval (47 meV, 50 meV) the opposite happens, the upper region with smaller twist angle $1.08^{\circ}$ is preferred. Similarly, in the TBG with intralayer hopping disorder, in the interval (37 meV, 42 meV) most states occupy the lower half (larger intralayer hopping equivalent to $1.20^{\circ}$) whereas states in (45 meV, 49 meV) live in the upper half (smaller intralayer hopping equivalent to $1.08^{\circ}$). Therefore, this phenomenology is a further reassurance of the very good qualitative agreement between both types of disorder.

It is worth mentioning that the scaling method `squeezes' the entire spectrum of TBG. It reduces the bandwidth of graphene from ($\Delta\nu\sim$ 18eV) to $\Delta\nu^{\prime}\sim\Delta\nu/\lambda$. The graphene van-Hove singularities thus appear at $\pm\Delta\nu^{\prime}/3$ and in principle the results are reliable in an interval (-$\Delta\nu^{\prime}/6$,+$\Delta\nu^{\prime}/6$). In practice, the scaling also leads to a slight blueshift of the spectrum and band narrowing. As a consequence, we find $\lambda\sim7$ to be the maximum practicable scaling factor for these exact diagonalization simulations. For the precise calculation of the transmission, which uses only a few unit cells, the requirement will be more stringent ($\lambda\lesssim4.5$), due to finite size effects.

\section*{S5. Decimation}
We develop a framework for the calculation of the local Green's function in a disordered region, within the path integral formulation of quantum mechanics and based on the technique of effective action. This allows us to compute the self-energies that are added to the disordered cell Green's function as a consequence of its coupling to the semi-infinite pristine leads, these self-energies are necessary for the calculation of the transmission through the disordered cell \cite{datta1997electronic}. The idea is to account for the effect of adjacent cells by a renormalization procedure which obtains frequency-dependent coupling matrices of the disordered cell to increasingly distant pristine cells, in the spirit of the decimation techniques presented in Refs. \cite{cea2019twists,guinea1983effective}.
We first show the derivation for a one dimensional chain of atoms, and then present the general matrix formulas for unit cells that contain multiple atoms. In the 1D case, the `disordered region' is a single impurity atom, the Hamiltonian of the full chain is

\begin{equation}
 H=\epsilon_0c_0^{\dagger}c_0 +t_R(c_0^{\dagger}c_1 +h.c.)+t_L(c_0^{\dagger}c_{-1} +h.c.)+\sum_{i=1}^{\infty}[\epsilon c_i^{\dagger}c_i+t(c_i^{\dagger}c_{i+1}+h.c.)] + \sum_{-\infty}^{-1}[\epsilon c_i^{\dagger}c_i+t(c_i^{\dagger}c_{i-1}+h.c.)] \, .
 \label{eq:S10}
\end{equation}

Here $\epsilon_0$ is the on-site energy of the impurity, $t_R$ ($t_L$) the hopping between the impurity and the atom to its right (left). $t$ is the hopping between atoms other than the impurity. Now consider the action of this system,

\begin{multline}
S=\int d\omega\Big[\sum_i\omega\overline{c}_i(\omega)c_i(\omega)-H\Big]=\int d\omega \Big[G_0^{-1}(\omega)\overline{c}_0(\omega)c_0(\omega)+\sum_{i\neq0}G^{-1}(\omega)\overline{c}_i(\omega)c_i(\omega)-t_R[\overline{c}_0(\omega)c_1(\omega)+\overline{c}_1(\omega)c_0(\omega)] \\
-t_L[\overline{c}_0(\omega)c_{-1}(\omega)+\overline{c}_{-1}(\omega)c_0(\omega)]-\sum_{i=1}^{\infty}t[\overline{c}_i(\omega)c_{i+1}(\omega)]+\overline{c}_{i+1}(\omega)c_i(\omega)]-\sum_{i=-\infty}^{-1}t[\overline{c}_i(\omega)c_{i-1}(\omega)]+\overline{c}_{i-1}(\omega)c_i(\omega)]\Big]\, ,
\label{eq:S11}
\end{multline}

where $c_i(\omega)$, $\overline{c}_i(\omega)$ are Grassmann numbers and $G_0(\omega)=\frac{1}{\omega I-\epsilon_{0}}$, $G(\omega)=\frac{1}{\omega I-\epsilon}$ are the local non-interacting Green's functions at the impurity and at a site $i\neq0$, respectively. Next, we integrate the variables $c_{\pm1}(\omega)$, $\overline{c}_{\pm1}(\omega)$ out of the action performing gaussian integration. This results in the effective action

\begin{multline}
S_{eff}^{(2)}=\int d\omega \Big[\big[G_0^{(2)}(\omega)\big]^{-1}\overline{c}_0(\omega)c_0(\omega)+\sum_{i=\pm 2}\big[G_i^{-1}(\omega)\overline{c}_i(\omega)c_i(\omega)-t_i(\omega)[\overline{c}_0(\omega)c_i(\omega)+\overline{c}_i(\omega)c_0(\omega)]\big]+ \\
+\sum_{\mid i\mid>2}G^{-1}(\omega)\overline{c}_i(\omega)c_i(\omega)-\sum_{i=2}^{\infty}t[\overline{c}_i(\omega)c_{i+1}(\omega)]+\overline{c}_{i+1}(\omega)c_i(\omega)]-\sum_{i=-\infty}^{-2}t[\overline{c}_i(\omega)c_{i-1}(\omega)]+\overline{c}_{i-1}(\omega)c_i(\omega)]\Big]\, ,
\label{eq:S12}
\end{multline}

where
\begin{equation}
\big[G_0^{(2)}(\omega)\big]^{-1}=G_0^{-1}(\omega)-t_R^2G(\omega)-t_L^2G(\omega)
\label{eq:S13}
\end{equation}

is the inverse of the local Green's function at the impurity,

\begin{equation}
G_{\pm 2}^{-1}(\omega)=G^{-1}(\omega)-t^2G(\omega)
\label{eq:S14}
\end{equation}

that at sites $\pm 2$ and

\begin{equation}
t_{\pm 2}(\omega)=tt_{R,L}G(\omega)
\label{eq:S15}
\end{equation}

are the effective frequency-dependent hopping amplitudes between the site 0 and sites $\pm 2$. We can iterate this procedure by integrating out all the degrees of freedom up to the sites $i=\pm(N-1)$. This leads to the effective action

\begin{multline}
S_{eff}^{(N)}=\int d\omega \Big[\big[G_0^{(N)}(\omega)\big]^{-1}\overline{c}_0(\omega)c_0(\omega)+\sum_{i=\pm N}\big[G_i^{-1}(\omega)\overline{c}_i(\omega)c_i(\omega)-t_i(\omega)[\overline{c}_0(\omega)c_i(\omega)+\overline{c}_i(\omega)c_0(\omega)]\big]+ \\
+\sum_{\mid i\mid>N}G^{-1}(\omega)\overline{c}_i(\omega)c_i(\omega)
-\sum_{i=N}^{\infty}t[\overline{c}_i(\omega)c_{i+1}(\omega)]+\overline{c}_{i+1}(\omega)c_i(\omega)]-\sum_{i=-\infty}^{-N}t[\overline{c}_i(\omega)c_{i-1}(\omega)]+\overline{c}_{i-1}(\omega)c_i(\omega)]\Big]\, .
\label{eq:S16}
\end{multline}

Eq. (\ref{eq:S16}) is analogous to (\ref{eq:S12}) and describes a system in which the impurity atom is coupled directly to sites $\pm N$ via frequency-dependent hopping amplitudes $t_{\pm N}$. The recursive formulas for the Green's functions $G_0^{N}$, $G_{\pm N}$ and the hopping functions $t_{\pm N}$ are,

\begin{equation}
\big[G_0^{(N)}(\omega)\big]^{-1}=\big[G_0^{N-1}(\omega)\big]^{-1}-t_{N-1}^2(\omega)G_{N-1}(\omega)-t_{-(N-1)}^2(\omega)G_{-(N-1)}(\omega)\, ,
\label{eq:S17}
\end{equation}

\begin{equation}
G_{\pm N}^{-1}(\omega)=G^{-1}(\omega)-t^2G_{\pm (N-1)}(\omega)\, ,
\label{eq:S18}
\end{equation}

\begin{equation}
t_{\pm N}(\omega)=tt_{\pm (N-1)}G_{\pm(N-1)}(\omega)\, .
\label{eq:S19}
\end{equation}

\begin{figure}[h]
\vspace*{+0.4cm}
    \centering
    {{\includegraphics[width=6.5cm]{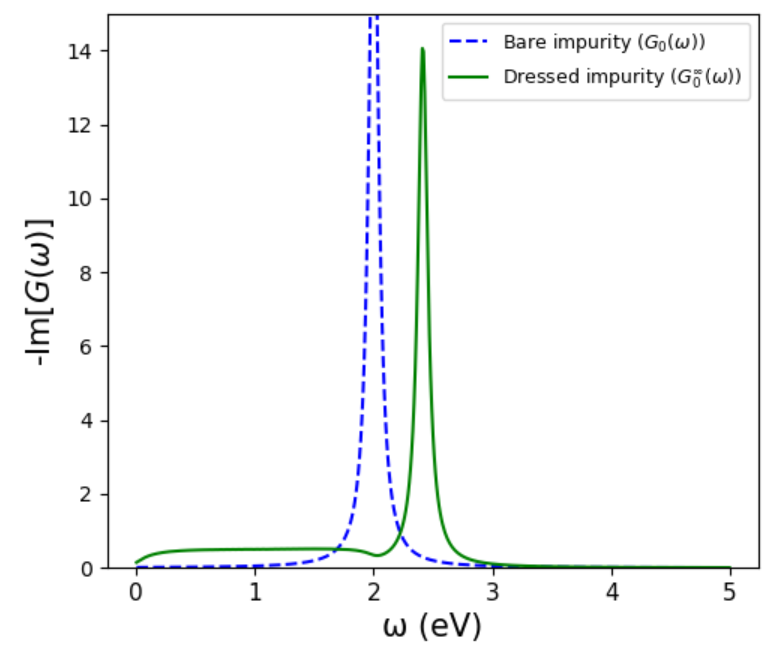} }}%
    \qquad
    {{\includegraphics[width=6.5cm]{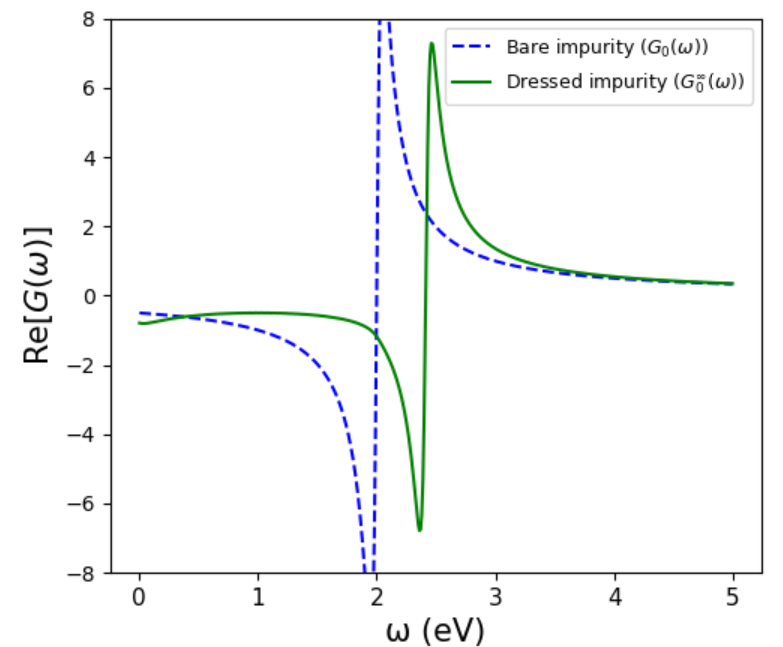} }}%
    \caption{Minus the imaginary part (left) and real part (right) of the impurity's retarded Green's function, before and after accounting for its coupling to the environment. Here $\epsilon_0=2$ eV, $\epsilon=1$ eV, $t=t_L=t_R=\frac{1}{2}$, $\delta=0.05$ eV. }%
    \label{fig:S11}
\end{figure}

Eqs. (\ref{eq:S13}-\ref{eq:S15}) set the step $N=2$ and the input to Eqs. (\ref{eq:S17}-\ref{eq:S19}). In the limit $N\rightarrow \infty$, $G_0^{(N)}$ converges to the local Green function at the impurity, including the effect of its environment. For practical purposes, one can compute the retarded Green's function, doing analytical projection on the upper half-plane, by substituting everywhere $\omega\rightarrow\omega+i\delta$, with $\delta$ a finite broadening. As an example, we consider an impurity with $\epsilon_0=2$ eV, the rest of atoms with $\epsilon=1$ eV, $t=t_L=t_R=\frac{1}{2}$ and $\delta=0.05$ eV. Figure \ref{fig:S11} shows the imaginary and real parts of $G_0^{(\infty)}$ for this set of parameters. Due to the environment, the impurity's Green function experiences a weak blueshift and damping at frequencies smaller than $\epsilon_0$.

\begin{figure}[h]
    \centering
    {{\includegraphics[width=12cm]{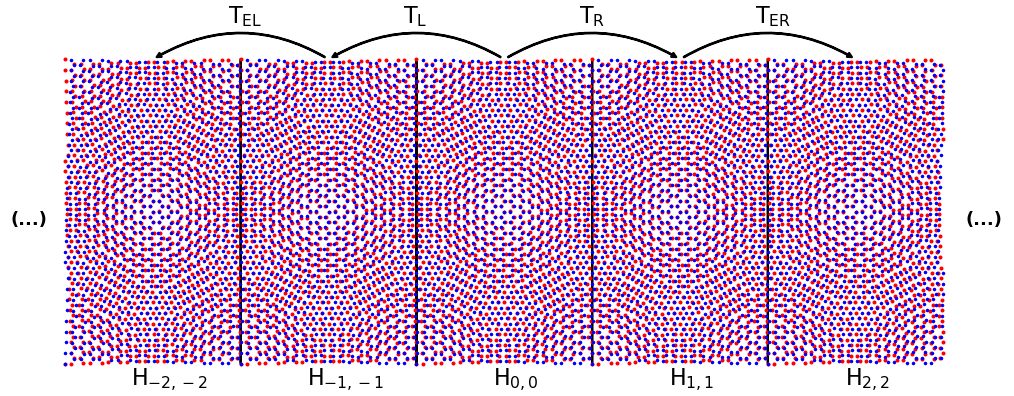} }}%
    \caption{Sketch of the geometry involved in the decimation technique. All cells to the left of the central one $H_{ii},i<0$ are pristine with uniform intralayer hopping $t^{\parallel}_1$. Cells to the right $H_{ii},i>0$ are pristine with a different hopping $t^{\parallel}_2$. These two semi-infinite leads are connected through the central disordered cell `0', whose Hamiltonian $H_{00}$ has intralayer hopping disorder ($t^{\parallel}_1$ evolving smoothly into $t^{\parallel}_2$ as an hyperbolic tangent). One needs to consider separately the coupling matrices from the disordered cell to its adjacent pristine cells ($T_L$ and $T_R$) and the coupling between pristine cells ($T_{EL}$ and $T_{ER}$). The goal is the calculation of the Green's function at the disordered cell, accounting for the effect of the semi-infinite pristine leads. Periodic boundary conditions are applied in the vertical direction.}%
    \label{fig:S12}
\end{figure} 

For multi-atomic cells, we start with the unperturbed Green's functions, $G_0(\omega)=\frac{1}{\omega I-H_{00}}$, $G(\omega)=\frac{1}{\omega I-H_{ii}}$, where $H_{00}$ is the Hamiltonian of the disordered cell and $H_{ii}$ the Hamiltonians of  pristine unit cells, equal for all $i>0$ and for all $i<0$. We also define $T_R=H_{0,1}$ and $T_{ER}=H_{i,i+1}$ with $i>0$ as well as $T_L=H_{0,-1}$ and $T_{EL}=H_{i,i-1}$ with $i<0$ (see Figure \ref{fig:S12}) the recursive formulas to obtain the Green's functions and coupling matrices are:

\begin{equation}
  \begin{aligned}
    \big[G_0^{(2)}(\omega)\big]^{-1}&=G_0^{-1}(\omega)-T_RG(\omega)T_R^{\dagger}-T_LG(\omega)T_L^{\dagger} \\        
    G_2^{-1}(\omega)&=G^{-1}(\omega)-T_{ER}^{\dagger}G(\omega)T_{ER}\\
    G_{-2}^{-1}(\omega)&=G^{-1}(\omega)-T_{EL}^{\dagger}G(\omega)T_{EL}\\
    T_{02}(\omega)&=T_RG(\omega)T_{ER}\\
    T_{20}(\omega)&=T_{ER}^{\dagger}G(\omega)T_R^{\dagger}\\
    T_{0,-2}(\omega)&=T_LG(\omega)T_{EL}\\
    T_{-2,0}(\omega)&=T_{EL}^{\dagger}G(\omega)T_L^{\dagger}\\ 
  \end{aligned}
\label{eq:S20}
\end{equation}

\begin{equation}
  \begin{aligned}
    \big[G_0^{(N+1)}(\omega)\big]^{-1}&=\big[G_0^{(N)}\big(\omega)]^{-1}-T_{0N}(\omega)G_N(\omega)T_{N0}(\omega)-T_{0,-N}(\omega)G_{-N}(\omega)T_{-N,0}(\omega) \\        
   G_{N+1}^{-1}(\omega)&=G^{-1}(\omega)-T_{ER}^{\dagger}G_N(\omega)T_{ER}\\
    G_{-N-1}^{-1}(\omega)&=G^{-1}(\omega)-T_{EL}^{\dagger}G_{-N}(\omega)T_{EL}\\
    T_{0,N+1}(\omega)&=T_{0N}(\omega)G_N(\omega)T_{ER}\\
    T_{N+1,0}(\omega)&=T_{ER}^{\dagger}G_N(\omega)T_{N0}(\omega)\\
    T_{0,-N-1}(\omega)&=T_{0,-N}(\omega)G_{-N}(\omega)T_{EL}\\
    T_{-N-1,0}(\omega)&=T_{EL}^{\dagger}G_{-N}(\omega)T_{-N,0}(\omega)\\ 
  \end{aligned}
\label{eq:S21}
\end{equation}

To obtain the transmission one also needs to compute advanced versions of all the quantities in (\ref{eq:S20}) and (\ref{eq:S21}), by doing analytical continuation with $\omega\rightarrow\omega-i\delta$ ($\delta$ a small broadening) instead of $\omega\rightarrow\omega+i\delta$ as in the retarded case. Once the matrices in (\ref{eq:S21}) have converged, the transmission can be readily calculated. First one computes the left and right `coupling functions' of the impurity cell to the semi-infinite leads \cite{datta1997electronic}:

\begin{equation}
\Gamma_{\{L,R\}}=i\big[\Sigma^r_{\{L,R\}}-\Sigma^a_{\{L,R\}}\big] \, .
\label{eq:S22}
\end{equation}

Here $\Sigma^r_{L}=\sum_{i}T_{0,-i}G_{-i}T_{-i,0}$ and $\Sigma^r_{R}=\sum_{i}T_{0,i}G_{i}T_{i,0}$ are the sums of all the terms subtracted from $[G_0(\omega)]^{-1}$ to reach the convergent value $[G^{\infty}_0(\omega)]^{-1}$, as a consequence of the coupling to the left and right pristine leads respectively. $\Sigma^a_{\{L,R\}}$ are the advanced versions. With these we can write the transmission,

\begin{equation}
\mathcal{T}=\textrm{Tr}\big\{\Gamma_LG_0^{\infty,r}\Gamma_RG_0^{\infty,a}\big\} \, .
\label{eq:S23}
\end{equation}

Note that $G^{\infty}_0(\omega)$ also gives us the local density of states at the disordered region,
\begin{equation}
\textrm{LDOS}=\frac{1}{\pi}\textrm{Tr}\big\{\textrm{Im}\{G^{\infty}_0(\omega)\}\big\} \, .
\label{eq:S24}
\end{equation}

Matrix inversion is the computational bottleneck of the algorithm, using standard BLAS libraries it scales with the size of the matrix as $N^2$. Initial simulations were attempted with TBG nanotube (19,1)@(-19,-1) (N=1016 and $\lambda\sim5$ for a magic angle simulation). Later we became aware that this system has finite size effects that lead to fine energy splittings of the flat bands. Therefore we recommend that calculations that demand high precision and employ one or few unit cells are done with scaling factor $\lambda\lesssim4.5$. However, exact diagonalization calculations that use many cells work well up to $\lambda\lesssim7$. The main results of the paper were obtained with (56,2)@(-56,-2), which like (28,1)@(-28,-1) has a scaling factor of $\lambda\sim 3.5$ and better resembles planar TBG since it fits two moir\'es in the direction perpendicular to the propagation.

\end{document}